\documentclass[preprint,showpacs,preprintnumbers,amsmath,amssymb,nofootinbib]{revtex4}
\usepackage[dvips]{graphicx}
\usepackage{graphicx}
\usepackage{dcolumn}
\usepackage{bm}
\usepackage{axodraw}

\newcommand{\bi}{\bibitem}
\newcommand{\be}{\begin{eqnarray}}
\newcommand{\ee}{\end{eqnarray}}
\newcommand{\bea}{\begin{array}}
\newcommand{\eea}{\end{array}}
\newcommand{\nn}{\nonumber}
\newcommand{\GeV}{\mbox{GeV}}
\catcode`\@=11
\def\lsim{\mathrel{\mathpalette\@versim<}}
\def\gsim{\mathrel{\mathpalette\@versim>}}
\def\@versim#1#2{\vcenter{\offinterlineskip
\ialign{$\m@th#1\hfil##\hfil$\crcr#2\crcr\sim\crcr } }}
\catcode`\@=12

\begin{document}
\def\descriptionlabel#1{\bf #1\hfill}
\def\description{\list{}{%
\labelwidth=\leftmargin
\advance \labelwidth by -\labelsep
\let \makelabel=\descriptionlabel}}

\vspace{2cm}

\title{R-Parity Violation and Non-Abelian Discrete Family Symmetry
\vspace{1cm}}

\author{Yuji Kajiyama\footnote{yuji.kajiyama@kbfi.ee}}
\affiliation{National Institute of Chemical Physics and Biophysics,
Ravala 10, Tallinn 10143, Estonia
\vspace{3cm}
}

\begin{abstract}
We investigate the implications of R-parity violating operators in a model 
with family symmetry.
The family symmetry can determine the form of R-parity violating operators 
as well as the Yukawa matrices responsible for fermion masses and mixings. 
In this paper we consider a concrete model with non-abelian discrete 
symmetry $Q_6$ which contains only three R-parity violating operators. 
We find that ratios of decay rates of the lepton flavor violating processes 
are fixed thanks to the family symmetry, predicting $BR(\tau \to 3e)/BR(\tau \to 3\mu)
\sim 4 m_{\mu}^2/m_{\tau}^2$.
\vspace{2cm}
\end{abstract}


\maketitle
\section{Introduction}
Despite the remarkable success of the gauge sector of the Standard Model (SM), there still exist some problems in the Higgs and Yukawa sectors. The Yukawa matrices are responsible for the masses and mixings of matter fermions: quarks and leptons. The Yukawa sector in the SM can give experimentally consistent masses and mixings, because it contains more free parameters than the number of observables in general. 
There is no predictivity in the Yukawa sector because of this redundancy of the parameters.
One of the ideas to overcome this issue is to introduce a family symmetry (flavor symmetry), which is the 
symmetry between generations. In this paper we consider a concrete model which is symmetric under the binary dihedral group $Q_6$ \cite{babukubo,kajiyama}. 
  
On the other hand, in the Higgs sector, the most important problem is that the Higgs boson has not been experimentally discovered yet. 
Discovery of the Higgs boson is expected at the Large Hadron Collider (LHC). 
In the SM, the Higgs mass 
is quadratic divergent. This problem is solved by introducing Supersymmetry (SUSY) 
at $O(1)$ TeV. 
The Minimal Supersymmetric Standard Model (MSSM) has low energy SUSY. 
In general it contains gauge symmetric, lepton and baryon number violating operators 
\be
W_{\not R}=\frac 12 \lambda_{ijk}L_i L_j E^c_k 
+\lambda'_{ijk}L_i Q_j D^c_k
+\frac 12 \lambda''_{ijk}U^c_i D^c_j D^c_k
+\mu_i H_u L_i
\label{rvmssm}
\ee
in addition to the usual Yukawa couplings and $\mu$-term. The asymmetric properties $\lambda_{ijk}=-\lambda_{jik}$ and $\lambda''_{ijk}=-\lambda''_{ikj}$ mean that $9+27+9+3=48$ (complex) parameters are included in this interactions. These couplings generate unacceptable processes such as   Lepton Flavor Violating (LFV) processes and proton decay. The conservation of R-parity 
\cite{farrar,rpreview} 
\be
R=(-1)^{3B+L+2s},
\ee
where $B$, $L$ and $s$ denote baryon, lepton number and spin of the particles, respectively, is one possibility to forbid the couplings Eq.(\ref{rvmssm}). From the definition, R-parity is $+1$ for all SM particles and $-1$ for all their superpartners. 
However, R-parity is not the only possible choice to forbid the interactions. 
Matter- or lepton- and baryon-parity \cite{ibanez} can be also a possibility. 
On the other hand, without R-parity,  these coupling constants have to be 
strictly constrained not to conflict with experimental data. 
Constraints on the R-parity violating couplings have been obtained by many authors 
from LFV processes \cite{hall,lfv,hinchliffe,gouvea,kim,chaichian}, neutrino mass 
\cite{hall,rnumass,joshipura,banks,bisset}, neutral meson system 
\cite{mesonmixing,carlos,bhattacharyya,saha,kundu,wang,grossman,jang,dreiner}, 
proton decay \cite{hinchliffe,pdecay,benhamo}, and so on \cite{rpreview,allanach,choudhury}. 

Family symmetries also constrain the form of R-parity interactions \cite{banks,benhamo,abelian,nonabelian} as well as the Yukawa matrices. 
In the model that we consider \cite{babukubo,kajiyama}, the $Q_6$ family symmetry reduces the 45 trilinear couplings to three: 
$\lambda$, $\lambda'_{1}$ and $\lambda'_{2}$. The baryon number violating couplings $\lambda''_{ijk}U^c_i D^c_j D^c_k$ are forbidden by the symmetry in our model, so it is guaranteed by the symmetry that the R-parity violating operators do not induce proton decay. 

In this paper, we study the phenomenology of  the three R-parity violating interactions in the model with 
$Q_6$ family symmetry \cite{babukubo,kajiyama}. First, we obtain upper bounds on three coupling constants 
$\lambda$ and $\lambda'_{1,2}$ from the experimental constraints. 
Next we focus on LFV processes induced by $\lambda$.  
The $\lambda LLE^c$ coupling generates the LFV decays 
$\ell_m^- \to \ell_i^- \ell_j^- \ell_k^+(m,i,j,k$ denote the flavor of the charged lepton) at tree level, and 
the Branching Ratios ($BR$) of the decay processes are proportional to $\lambda^4$.
Therefore, the ratios of these processes are independent of $\lambda$ and can be predicted 
unambiguously to be  
$BR(\tau \to eee)/BR(\tau \to \mu\mu\mu)\sim~4m_{\mu}^2/m_{\tau}^2$. 
It reflects the properties of the family symmetry. 
We introduce the $Q_6$ symmetric model in the next section, and derive the predictions in the sect. III.

\section{The Model}
\subsection{Group Theory of $Q_6$}
The binary dihedral group $Q_N(N=2,4,6,...)$ is a finite subgroup of $SU(2)$ and defined by the following set of $2N$ elements
\be
Q_N&=&\left\{ 1,A,A^2,...,A^{N-1},B,AB,...,A^{N-1}B\right\} ,
\ee
where two dimensional representation of matrix $A$ and $B$ is given by
\be
A&=&\left(\bea{cc} 
\cos \phi_N & \sin \phi_N \\
-\sin \phi_N & \cos \phi_N \\
\eea \right),~\phi_N=\frac{2 \pi}{N},~~~
B=\left( \bea{cc} 
i & \\
 & -i \\
\eea \right).
\label{QNdef}
\ee
Since we consider a supersymmetric model with $Q_6$ family symmetry,
 we show the multiplication rules only for the case of $N=6$.

$Q_6$ group contains 2 two-dimensional irreducible representations (irreps), 
${\bf 2}_1,{\bf 2}_2$ and 4 one-dimensional ones ${\bf 1}_{+,0},{\bf 1}_{+,2},{\bf 1}_{-,1},{\bf 1}_{-,3}$, where ${\bf 2}_1$ is pseudo real and ${\bf 2}_2$ is real representation. 
In the notation of ${\bf 1}_{\pm,n}(n=0,1,2,3)$, $\pm$ stands for the change of sign under the transformation by matrix $A$, and $n$ the factor $\exp(i n \pi/2)$ by $B$. So ${\bf 1}_{+,0}$ and ${\bf 1}_{+,2}$ are 
real representations, while ${\bf 1}_{-,1}$ and ${\bf 1}_{-,3}$ are complex conjugate to each other. 
Their group multiplication rules are given as follows \cite{babukubo,frampton,kajiyama}: 
\be
{\bf 1}_{+,2} \times {\bf 1}_{+,2}={\bf 1}_{+,0},~
{\bf 1}_{-,3} \times {\bf 1}_{-,3}&=&{\bf 1}_{+,2},~
{\bf 1}_{-,1} \times {\bf 1}_{-,1}={\bf 1}_{+,2},~
{\bf 1}_{-,1} \times {\bf 1}_{-,3}={\bf 1}_{+,0},~ \nn \\
{\bf 1}_{+,2} \times {\bf 1}_{-,1}={\bf 1}_{-,3},~
{\bf 1}_{+,2} \times {\bf 1}_{-,3}&=&{\bf 1}_{-,1},~~~~
{\bf 2}_{1} \times {\bf 1}_{+,2}={\bf 2}_{1},~~~~~
{\bf 2}_{1} \times {\bf 1}_{-,3}={\bf 2}_{2},~\nn \\
{\bf 2}_{1} \times {\bf 1}_{-,1}={\bf 2}_{2},~~~~~~
{\bf 2}_{2} \times {\bf 1}_{+,2}&=&{\bf 2}_{2},~~~~~~
{\bf 2}_{2} \times {\bf 1}_{-,3}={\bf 2}_{1},~~~~~~
{\bf 2}_{2} \times {\bf 1}_{-,1}={\bf 2}_{1},
\label{multi1}
\ee

\be
\begin{array}{ccccccccc}
 {\bf 2}_1  &  \times   
&  {\bf 2}_1  &  =  &  {\bf 1}_{+,0} 
&  +   &  {\bf 1}_{+,2} & + &  {\bf 2}_2 
\\ 
 \left(\begin{array}{c}x_1 \\ x_2  \end{array} \right)   & 
 \times    &  \left(\begin{array}{c}y_1 \\  y_2   \end{array}\right)  
&  =  &   (x_1 y_2 - x_2 y_1)   &  &
 (x_1 y_1 + x_2 y_2)   &    &
 \left(\begin{array}{c}-x_1 y_2 - x_2 y_1 \\ x_1 y_1 - x_2 y_2 \end{array} \right) ,\\ 
\end{array}
\label{multi2} 
\ee
\be
\begin{array}{ccccccccc}
 {\bf 2}_2  &  \times   
&  {\bf 2}_2  &  =  &  {\bf 1}_{+,0} 
&  +   &  {\bf 1}_{+,2} & + &  {\bf 2}_2 
\\ 
 \left(\begin{array}{c}a_1 \\ a_2  \end{array} \right)   & 
 \times    &  \left(\begin{array}{c}b_1 \\  b_2   \end{array}\right)  
&  =  &   (a_1 b_1 + a_2 b_2)   &  &
 (a_1 b_2 -a_2 b_1)   &    &
 \left(\begin{array}{c}-a_1 b_1 + a_2 b_2 \\ a_1 b_2 +a_2 b_1 \end{array} \right) ,\\ 
\end{array}
\label{multi3} 
\ee
\be
\begin{array}{ccccccccc}
 {\bf 2}_1  &  \times   
&  {\bf 2}_2  &  =  &  {\bf 1}_{-,3}
&  +   &  {\bf 1}_{-,1} & + &  {\bf 2}_1 
\\ 
 \left(\begin{array}{c}x_1 \\ x_2  \end{array} \right)   & 
 \times    &  \left(\begin{array}{c}a_1 \\  a_2 \end{array} \right)  
&  =  &   (x_1 a_2 + x_2 a_1)   &  &
 (x_1 a_1 -x_2 a_2)   &    &
 \left(\begin{array}{c}x_1 a_1 + x_2 a_2 \\ x_1 a_2 -x_2 a_1 \end{array} \right) .  \\ 
\end{array}
  \label{multi4}
\ee
In what follows we construct a concrete model with $Q_6$ family symmetry 
by using of the multiplication rules above.

\subsection{$Q_{6}$ assignment and superpotential }
We show the $Q_{6}$ assignment of the quark, 
lepton and Higgs chiral supermultiplets below,
 where
$Q_I, Q_3, L_I, L_3$ and  $ H_I^u, H_3^u, H_I^d, H_3^d$
stand for $SU(2)_L$ doublets
supermultiplets for quarks, leptons and Higgs bosons, respectively.
Similarly, $SU(2)_L$ singlet
supermultiplets for quarks, charged leptons and neutrinos are denoted by
$U_I^c, U^c_3,D_I^c, D^c_3, E_I^c, E^c_3$ and $N_I^c, N^c_3$. 
The generation indices $I,J,...=(1,2)$ are applied to the $Q_6$ doublet, and 
$i,j,...=(1,2,3)$ to three generations throughout the paper. 
$Y$ is gauge singlet Higgs supermultiplet to give neutrino mass by seesaw mechanism.

We give the $Q_6$ assignment to each field as follows:
 \be
{\bf 2}_1~~&:&~~Q_I, \nn \\
{\bf 2}_2~~&:&~~U_I^c,~D_I^c,~\hat L_I,~E_I^c,~N_I^c,~H^u_I,~\hat H^d_I,\nn \\
{\bf 1}_{+,0}~~&:&~~ L_3,~E_3^c,\nn \\
{\bf 1}_{+,2}~~&:&~~Q_3,~Y, \label{assignment2} \\
{\bf 1}_{-,1}~~&:&~~U_3^c,~D_3^c,~H^u_3,~H^d_3, \nn \\
{\bf 1}_{-,3}~~&:&~~N_3^c. \nn
\ee
In a model without R-parity conservation, there is no distinction between lepton doublet 
and down type Higgs doublet, because both have the same gauge quantum numbers. 
We have written these fields as $\hat L_I,\hat H^d_I$, 
and physical lepton doublet and 
down type Higgs doublet will be written as linear combination of these.
At first we write down the superpotential in the fields with hat ($\hat L_I,\hat H^d_I$),
 and after that we rewrite it 
in the physical fields without hat ($ L_I,H^d_I$).
    
Under the field assignment above, we can write down the most general, renormalizable  
$Q_6$ invariant superpotential $W$ (without R-parity conservation):
\be
W=\hat W_Q+\hat W_L+\hat W_{\mu}+\hat W_{R\hspace{-2mm}/},
\label{ww}
\ee   
where 
 \be
\hat W_Q&=&W_U+\hat W_D, \nn \\ 
W_U&=&Y^u_a Q_3 U^c_3H^u_3+Y^u_bQ_I \left( \sigma^1 \right)_{IJ}U^c_3 H^u_J
-Y^u_{b'}Q_3 U^c_I \left( i\sigma^2 \right)_{IJ}H^u_J
+Y^u_c Q_I \left( \sigma^1 \right)_{IJ}U^c_J H^u_3, \nn \\
\hat W_D&=&Y^d_a Q_3 D^c_3H^d_3+\hat Y^d_bQ_I \left( \sigma^1 \right)_{IJ}D^c_3 \hat H^d_J
-\hat Y^d_{b'}Q_3 D^c_I \left( i\sigma^2 \right)_{IJ}\hat H^d_J
+Y^d_c Q_I \left( \sigma^1 \right)_{IJ}D^c_J H^d_3 \nn \\
&~&
\label{wq}
\ee
for the quark sector \footnote{The superpotential for the up quark sector $W_U$ does not have 
hat ($\hat~$), because it contains neither $\hat L_I$ nor $\hat H^d_I$.}, and 
\be
\hat W_L&=&\hat W_E+\hat W_N, \nn \\
\hat W_E&=&\hat Y_b^e \hat L_I E_3^c \hat H_I^d+\hat Y_{b'}^e L_3 E_I^c \hat H_I^d
+\hat Y_c^e f_{IJK} \hat L_I E_J^c \hat H_K^d, \nn \\
\hat W_N&=&Y_a^{\nu}L_3 N_3^c H_3^u+Y_{b'}^{\nu}L_3 N_I^c H_I^u 
+\hat Y_c^{\nu}f_{IJK}\hat L_I N_J^c H_K^u
+\frac{1}{2} M_N N_I^c N_I^c +\lambda_N Y N^c_3 N_3^c, \nn \\
&~&-f_{111}=f_{221}=f_{122}=f_{212}=1
\label{wl}
\ee
for the lepton sector.
The R-parity violating trilinear coupling $W_{R\hspace{-2mm}/}$ which respects the 
family symmetry is given by
\be
\hat W_{R\hspace{-2mm}/}=\hat \lambda L_3 \hat L_I E_I^c
+\hat \lambda_1'  \hat L_I (i \sigma^2)_{IJ}Q_3 D_J^c
+\hat \lambda_2'  \hat L_I (\sigma^1)_{IJ}Q_J D_3^c.
\label{wr}
\ee
The family symmetry constrains not only the form of the Yukawa sector but also that of the 
R-parity violating couplings. 
Proton decay caused by dimension-four operators 
$\lambda''_{ijk}U^c_i D^c_j D^c_k$ are prevented
 since the baryon number violating interactions 
are forbidden by the symmetry \cite{kajiyama}.
Therefore R-parity violating trilinear terms 
$\hat W_{R\hspace{-2mm}/}$ contain only three couplings $\hat \lambda$ and 
$\hat \lambda'_{1,2}$. 
It should be compared with the MSSM case in which it contains 45 (complex) trilinear couplings.
We will analyze the phenomenology of the couplings Eq. (\ref{wr}) in the section III.

In the following analysis, 
we assume that any couplings appearing in the superpotential Eq.(\ref{ww}) are real, and 
CP symmetry is violated spontaneously by vacuum expectation values (VEVs) of 
the Higgs bosons.  
The $\mu$-term $W_{\mu}$ which includes both R-parity conserving and violating bilinear terms 
is discussed in the next subsection.

\subsection{Bilinear terms}
In this subsection, we discuss bilinear terms and define the physical fields $L_I$ and $H^d_I$.
Since the lepton and down type Higgs doublet have the same gauge quantum numbers, 
it is natural that both $\hat L_I$ and $\hat H^d_I$ get VEVs. 
We assume that the fields get complex VEVs 
\be
\langle H^u_1\rangle=\langle H^u_2\rangle&=&\frac 12 v_D^u e^{i \theta^u},~
\langle H^u_3 \rangle=\frac{1}{\sqrt{2}}v_3^u e^{i \theta^u_3},~ 
\langle H^d_3 \rangle=\frac{1}{\sqrt{2}}v_3^d e^{i \theta^d_3}, 
\label{vev1}\\
\langle \hat H^d_1\rangle=\langle \hat H^d_2 \rangle &=& \frac 12 v_d e^{i \theta^d},~
\langle \hat L_1\rangle=\langle \hat L_2 \rangle = \frac 12 v_L e^{i \theta^d},
\label{vev2}
\ee
where the VEVs of $\hat H^d_I$ and $\hat L_I$ have the same phase. 
Also, the $Q_6$ singlet lepton doublet $L_3$ does not have non-zero VEV by definition.
To ensure this VEV structure, the $\mu$-term should have the symmetry
\be
H^u_1 \leftrightarrow H^u_2,~\hat H^d_1 \leftrightarrow \hat H^d_2,
~\hat L_1 \leftrightarrow \hat L_2, 
\label{z2}
\ee
and it is written as \footnote{This form of the $\mu$-term is given by introducing 
additional gauge singlet Higgs fields and extra discrete symmetry \cite{babukubo, kajiyama}. 
In the present paper, we just assume the form of Eq.(\ref{muterm}).} 
\be 
\hat W_{\mu}&=&\mu^d_1 H^u_I \hat H^d_I +\mu^d_3 H^u_3 H^d_3 +\mu^d_{13}(H^u_1+H^u_2)H^d_3
+\mu^d_{31}H^u_3(\hat H^d_1+\hat H^d_2)+\mu^d_{12}H^u_I (\sigma^1)_{IJ}\hat H^d_J\nn \\
&+&\mu^L_1 H^u_I \hat L_I +\mu^L_{31}H^u_3(\hat L_1 +\hat L_2)
+\mu^L_{12}H^u_I (\sigma^1)_{IJ}\hat L_J.
\label{muterm}
\ee
We define the physical Higgs fields as the fields whose VEVs break 
$SU(2)_L \times U(1)_Y$ symmetry. So the physical Higgs fields $H^d_I$ and 
lepton doublets $L_I$ are defined by the linear combination
\be
H^d_I \equiv \frac{v_d}{v_D^d}\hat H^d_I+\frac{v_L}{v_D^d}\hat L_I,~
L_I \equiv \frac{v_L}{v_D^d}\hat H^d_I-\frac{v_{d}}{v_D^d}\hat L_I, 
\label{tildehl}
\ee  
where $v_D^d=\sqrt{v_d^2+v_L^2}$. Therefore VEVs of $H^d_I$ are 
\be
\langle H^d_I \rangle=\frac{v_D^d}{2} e^{i \theta^d},
\label{vev3}
\ee
which break the electroweak symmetry, while $L_I$ do not 
get VEVs. 

The $\mu$-term Eq.(\ref{muterm}) should be rewritten in the new Higgs fields $H^d_I$ and 
lepton doublet $L_I$ as follows:
\be
W_{\mu}&=& \mu_1 \cos \xi_1 H^u_I H^d_I 
+ \mu^d_3 H^u_3 H^d_3+\mu^d_{13}(H^u_1+H^u_2)H^d_3
+\mu_{31}\cos \xi_{31}H^u_3 (H^d_1+H^d_2)
+\mu_{12} \cos \xi_{12}H^u_I (\sigma^1)_{IJ}H^d_J \nn \\
&+&\mu_1 \sin \xi_1 H^u_I L_I
+\mu_{31} \sin \xi_{31}H^u_3(L_1+L_2)
+\mu_{12} \sin \xi_{12}H^u_I(\sigma^1)_{IJ}L_J,
\ee
where the mixing angles of the Higgs fields and lepton doublets are defined as  
\be
\mu_1 \sin \xi_1=\frac{\mu^d_1 v_L-\mu^L_1 v_d}{v_D^d},~\mbox{etc.} 
\ee
with $\mu_1=\sqrt{(\mu^d_1)^2+(\mu^L_1)^2}$ etc.

These mixing terms of neutralinos and neutrinos generate neutrino masses proportional to 
$\sin^2 \xi$ at tree level \cite{joshipura,banks}.

There are particularly strong constraints on the mixing angle $\sin \xi$ coming from neutrino masses, 
and it means that the $\mu$-terms and VEVs have to be aligned, 
$\mu^d_1/\mu^L_1 \propto v_d/v_L$\footnote{In Ref \cite{banks}, possibilities of the alignment 
have been discussed in the framework of $U(1)$ horizontal symmetries.}.
In the MSSM case, a neutrino mass bound requires a strong alignment of 
the mixing term $W \sim \mu \sin \xi H_u L_3$, 
\be
\sin \xi < 3\times 10^{-6} \sqrt{1 +\tan^2 \beta}.
\label{sinxi}
\ee
Therefore we simply neglect the mixing terms in the following analysis. 

We can rewrite the superpotential Eqs.(\ref{ww})$\sim$(\ref{wr}) in the completely same form 
with the physical fields, with new coupling constants defined as
\be
Y^d_a&=&\hat Y^d_a,~Y^d_b=\frac{v_d}{v_D^d}\hat Y^d_b-\frac{v_L}{v_D^d}\hat \lambda'_2,~
Y^d_{b'}=\frac{v_d}{v_D^d}\hat Y^d_{b'}-\frac{v_L}{v_D^d}\hat \lambda'_1,~Y^d_c=\hat Y^d_c, \nn \\
Y^e_b&=&-\hat Y^e_b,~
Y^e_{b'}=\frac{v_d}{v_D^d}\hat Y^e_{b'}+\frac{v_L}{v_D^d}\hat \lambda,~
Y^e_c=-\hat Y^e_c,  \label{newY} \\
\lambda&=&\frac{v_L}{v_D^d}\hat Y^e_{b'}-\frac{v_d}{v_D^d}\hat \lambda,~
\lambda'_1=-\frac{v_L}{v_D^d}\hat Y^d_{b'}-\frac{v_d}{v_D^d}\hat \lambda'_1,~
\lambda'_2=-\frac{v_L}{v_D^d}\hat Y^d_b-\frac{v_d}{v_D^d}\hat \lambda'_2. \nn 
\ee  
In what follows, we will discuss the phenomenology of these new superpotential 
written in unhatted fields.
 
\subsection{Fermion mass matrices and 
diagonalization}
 We assume that VEVs take the form Eq.(\ref{vev1}) and (\ref{vev3}), from which
we obtain the fermion mass matrices.
 
 \subsubsection{Quark sector}
 The quark mass matrices are given by
\be
{\bf m}^{u} &=&\frac{1}{2}\left(\begin{array}{ccc}
0 &\sqrt{2} Y_c^{u} v_3^u e^{-i \theta^u_3}
& Y_b^{u}  v_D^u e^{-i \theta^u}\\
\sqrt{2} Y_c^{u}v_3^u e^{-i \theta^u_3} & 0 &
 Y_b^{u} v_D^ue^{-i \theta^u} \\
-Y_{b'}^{u} v_D^ue^{-i \theta^u} & 
Y_{b'}^{u}v_D^u e^{-i \theta^u} & 
\sqrt{2} Y_a^{u}  v_3^u e^{-i \theta^u_3} \\
\end{array}\right),
\label{mu}\\
{\bf m}^{d} &=&\frac{1}{2}\left(\begin{array}{ccc}
0 & \sqrt{2}Y_c^{d} v_3^d e^{-i \theta^d_3}
&Y_b^{d}  v_D^d e^{-i \theta^d}\\
\sqrt{2} Y_c^{d}v_3^d e^{-i \theta^d_3} & 0 &
Y_b^{d} v_D^de^{-i \theta^d} \\
-Y_{b'}^{d} v_D^de^{-i \theta^d} & 
 Y_{b'}^{d}v_D^d e^{-i \theta^d} & 
\sqrt{2}Y_a^{d}  v_3^d e^{-i \theta^d_3} \\
\end{array}\right).
\label{md}
\ee

We can bring the mass matrices above to the form
\be
{\bf \hat{m}}^u = m_t\left(\begin{array}{ccc}
0 & q_u/y_u & 0  \\ -q_u/y_u & 0 & b_u\\
0 &   b_u'  &  y^2_u \\
\end{array}\right),
\label{muhat}
\ee
by $\pi/4$ rotations on the $Q_6$ doublet fermions and appropriate phase rotations.
This mass matrix can be then diagonalized by orthogonal matrices $O^u_{L,R}$ as
\footnote{The form of the mass matrix is known as
the next-neighbor interaction form \cite{weinberg,fritzsch1}.}
\be
O^{uT}_L {\bf \hat{m}}^u O_R^u &=&
\left( \begin{array}{ccc}m_u & 0 & \\ 0 & m_c & 0\\ 0 & 0 & m_t
\end{array}\right), 
\ee
and similarly for ${\bf m}^d$.

For the set of the parameters
 \be
\theta_q & = &\theta^d_3-\theta^d-\theta^u_3
+\theta^u  =-1.25,
q_u=0.0002150,
b_u=0.04440,
b'_u=0.09300,\nn\\
y_u &=&0.99741,
q_d =0.005040,
b_d=0.02500,
b'_d=0.7781,
y_d=0.7970,
\label{parameters}
\ee
we obtain
\be
m_u/m_t &=& 1.11\times 10^{-5},
m_c/m_t=4.14 \times 10^{-3},
m_d/m_b=1.22 \times 10^{-3},
m_s/m_b=2.07 \times 10^{-2},
\nn\\
|V_{\rm CKM}| &=&
\left( \begin{array}{ccc}
0.97440 & 0.2267  & 0.00388
\\  0.2265   & 0.9731& 0.0421
  \\ 0.00924 &0.0412& 0.9991
\end{array}\right),~
\sin 2\beta (\phi_1)=0.723.
\label{predQ}
\ee
These values are consistent with present experimental values \cite{pdg}.  
So, we see that the model can well reproduce the experimentally measured 
parameters.
Moreover, since the CKM parameters and the quark masses are related to 
each other because of the 
family symmetry, we find that {\em nine} independent parameters (Eq.(\ref{parameters})) of the model 
can well describe {\em ten} physical observables: there is {\em one} prediction. 
An example of the prediction is $|V_{td}/V_{ts}|$,
whose experimental value has been obtained from
the measurement
of  the mass difference $\Delta m_{B_s}$ of the $B_s^0$ meson \cite{deltams}:
\be
\mbox{Model}&:& 
|V_{td}/V_{ts}| =0.21-0.23,\nn\\
\mbox{Exp.} &:& |V_{td}/V_{ts}|
=0.208\begin{array}{c}+0.001\\-0.002\end{array}
\mbox{(exp.)}
\begin{array}{c}+0.008\\-0.006\end{array}
\mbox{(theo.)}.
\label{pred3}
\ee

Finally, we give the unitary matrices that rotate the quarks
for the choice of the parameters given in Eq.(\ref{parameters}):
\be
U_{uL}&=&\left(\begin{array}{ccc}
0.706  & 0.0366 &  1.42 \times 10^{-5} \\
-0.706  & -0.0366  & -1.42 \times 10^{-5} \\
0 & 0 &  0\\
\end{array}\right), \nn \\
&+&e^{2 i \Delta \theta^u}\left(\begin{array}{ccc}
0.0366 & -0.705 & 0.0313 \\
0.0366 & -0.705 & 0.0313 \\
-0.00231 ~e^{-i \Delta \theta^u} & 0.0441~ e^{-i \Delta \theta^u} & 
0.999~ e^{-i \Delta \theta^u} \\
\end{array}\right), \label{UuL}
\ee
\be
U_{dL}&=&\left(\begin{array}{ccc}
0.695  & 0.130 &  0.00345 \\
-0.695  & -0.130  & -0.00345 \\
0 & 0 &  0\\
\end{array}\right), \nn \\ 
&+&e^{2 i \Delta \theta^d}\left(\begin{array}{ccc}
0.130 & -0.695 & 0.0111 \\
0.130 & -0.695 & 0.0111 \\
-0.00769 ~e^{-i \Delta \theta^d} & 0.0146~ e^{-i \Delta \theta^d} & 
1.00~ e^{-i \Delta \theta^d} \\
\end{array}\right), \label{UdL} 
\ee
\be 
U_{uR}&=&e^{i \theta^u_3}\left(\begin{array}{ccc}
0.0366  & 0.703 &  0.0657 \\
0.0366  & 0.703  & 0.0657 \\
0 & 0 &  0\\
\end{array}\right)  \nn \\ 
&+&e^{2 i \Delta \theta^u+i \theta^u_3}\left(\begin{array}{ccc}
-0.706 & 0.0367 & -6.73 \times 10^{-6} \\
0.706 & -0.0367 & 6.74  \times 10^{-6} \\
-0.00484 ~e^{-i \Delta \theta^u} & -0.0928~ e^{-i \Delta \theta^u} & 
0.996~ e^{-i \Delta \theta^u} \\
\end{array}\right), \label{UuR} 
\ee
\be
U_{dR}&=&e^{i \theta^d_3}\left(\begin{array}{ccc}
0.134  & 0.427 &  0.548 \\
0.134  & 0.427  & 0.548 \\
0 & 0 &  0\\
\end{array}\right)  \nn \\ 
&+&e^{2 i \Delta \theta^d+i \theta^d_3}\left(\begin{array}{ccc}
-0.675 & 0.212 & -7.01 \times 10^{-5} \\
0.675 &- 0.212 & 7.01 \times 10^{-5} \\
-0.232 ~e^{-i \Delta \theta^d} &- 0.739~ e^{-i \Delta \theta^d} & 
0.633~ e^{-i \Delta \theta^d} \\
\end{array}\right). \label{UdR}
\ee
The unitary matrices above will be used 
when discussing R-parity violating processes in sect. III.
\subsubsection{Lepton sector}
The mass matrix in the charged lepton sector is:
\be
{\bf m}^{e} &=&\frac{1}{2} \left(\begin{array}{ccc}
- Y_{c}^{e}& 
Y_{c}^{e}
&  Y_b^{e} \\
 Y_{c}^{e} & 
  Y_{c}^{e} &
 Y_b^{e}  \\
Y_{b'}^{e}  & 
Y_{b'}^{e} & 
0 \\
\end{array}\right)  v_D^d e^{-i \theta^d}.
\label{me}
\ee
It is diagonalized by the biunitary transformation:
\be
U_{eL}^{\dag}{\bf m}^{e} U_{eR} &=&
\left( \begin{array}{ccc}m_e & 0 & \\ 0 &
 m_\mu & 0\\ 0 & 0 & m_\tau
\end{array}\right).
\ee
One finds that  $U_{eL}$ and $ U_{eR}$ can be approximately 
written
as
\be
U_{eL}&=&\left( \bea{ccc} 
\epsilon_e (1-\epsilon_{\mu}^2)& (1/\sqrt{2})
(1-\epsilon_e^2+\epsilon_e^2 \epsilon_{\mu}^2)&1/\sqrt{2} \\
-\epsilon_e (1+\epsilon_{\mu}^2)& -(1/\sqrt{2})
(1-\epsilon_e^2-\epsilon_e^2 \epsilon_{\mu}^2)&1/\sqrt{2} \\
1-\epsilon_e^2 & -\sqrt{2} \epsilon_e & \sqrt{2}\epsilon_e \epsilon_{\mu}^2 \\
\eea \right), \label{UeL} \\
U_{eR}&=& \left( \bea{ccc} 
-\epsilon_e^2 (1-\epsilon_{\mu}^2/2)&-1&0 \\
1-\epsilon_{\mu}^2/2 & -\epsilon_e^2 (1-\epsilon_{\mu}^2) &\epsilon_{\mu} \\
-\epsilon_{\mu} & \epsilon_e^2 \epsilon_{\mu} & 1-\epsilon_{\mu}^2/2 \\
\eea \right)e^{i \theta^d}, \label{UeR}
\ee
and small parameters $\epsilon_e,\epsilon_{\mu}$ are defined as
\be
\epsilon_e=\frac{m_e}{\sqrt{2}m_{\mu}}=3.42 \times 10^{-3},~
\epsilon_{\mu}=\frac{m_{\mu}}{m_{\tau}}=5.94 \times 10^{-2}.
\label{epsilon}
\ee
In the limit $m_e=0$, the unitary matrix $U_{eL}$
becomes
$$
\left(
\begin{array}{ccc}
0&1/\sqrt{2} &1/\sqrt{2}\\
0 &-1/\sqrt{2}  &1/\sqrt{2}\\1& 0 &  0
\end{array}\right),
$$
which is the origin for a maximal mixing of
the atmospheric neutrinos.

As for the neutrino sector, we assume that
a see-saw mechanism \cite{seesaw} takes place.
However, we do not present the details of the neutrino sector here because 
there is no need to know it in the following analysis. We obtain some specific predictions of our model:
(i) only an inverted mass hierarchy $m_{\nu_3}<m_{\nu_1},m_{\nu_2}$ is consistent with 
the experimental constraint $|\Delta m_{21}^2|< |\Delta m_{23}^2|$, 
(ii) the $(e,3)$ element of the MNS matrix is given by $|U_{e3}| \simeq \epsilon_e$. 
See Refs. \cite{kubo,kajiyama} for details.

\subsection{Soft supersymmetry breaking sector}
Next we consider the soft SUSY breaking sector, which respects the 
family symmetry. 
If three generations of a family is put into
a one-dimensional and two-dimensional irreps of 
any dihedral group, then 
the soft scalar mass matrix for sfermions has always a diagonal form: 
\be
{\bf {m}^2}_{(\tilde q,\tilde \ell)LL} =
\left(
\begin{array}{ccc}
{m}^2_{(\tilde q,\tilde \ell)1L}  & 0 & 0 \\
0 & {m}^2_{(\tilde q,\tilde \ell)1L}& 0 \\
0 & 0 & {m}^2_{(\tilde q,\tilde \ell)3L} \\
\end{array}
\right),~
{\bf {m}^2}_{\tilde aRR} =
\left(
\begin{array}{ccc}
{m}^2_{\tilde a1R}  & 0 & 0 \\
0 & {m}^2_{\tilde a1R}  & 0 \\
0 & 0 & {m}^2_{\tilde a3R} \\
\end{array}
\right)~~~(a=u,d,e).
\label{scalarmass}
\ee

Note that the mass of the first two generations are degenerated because of the family symmetry.
\footnote{SUSY flavor and CP problem can be avoided thanks to such partially 
degenerated (Eq.(\ref{scalarmass})) and aligned (Eq.(\ref{Aterm})) soft terms because of 
the family symmetry\cite{babukubo,kajiyama}. } 
Further, since the trilinear  interactions ($A$-terms)
are also  $Q_6$ invariant,
the left-right mass matrices have the form
\be
\left({\bf {m}}^2_{\tilde aLR}\right)_{ij} 
&=&
A_{ij}^a\left( {\bf m}^a \right)_{ij} 
~~(a=u,d,e),
\label{Aterm}
\ee
where $A_i^{a}$'s are free parameters of dimension one,
and the fermion masses ${\bf m}$'s are given in
Eq.(\ref{mu}), (\ref{md}) and (\ref{me}).
They are real, because we impose CP invariance at
the Lagrangian level.

We approximate that the squark and slepton masses are 
given only by Eq.(\ref{scalarmass}), that is, trilinear terms Eq.(\ref{Aterm}) are 
negligible \cite{kajiyama}. 

\section{R-Parity Violation}
Since $Q_6$ family symmetry controls the whole flavor structure of the model, the form of the R-parity violating couplings are also constrained by the family symmetry. We find that only possible trilinear couplings 
allowed by the symmetry can be written as
\be
W_{R\hspace{-2mm}/}=\lambda L_3 L_I E_I^c+\lambda_1'  L_I (i \sigma^2)_{IJ}Q_3 D_J^c
+\lambda_2'  L_I (\sigma^1)_{IJ}Q_J D_3^c
\label{rparity}
\ee
in the physical lepton doublet $L_I$ defined in  Eq.(\ref{tildehl}) with the coupling constants 
in Eq.(\ref{newY}). 
Here the superpotential $W_{R\hspace{-2mm}/}$ is written in the flavor eigenstates, so 
the mixing matrices Eqs.(\ref{UuL})$\sim$(\ref{UdR}), (\ref{UeL}) and (\ref{UeR}) 
should appear when we rotate the fermion components into their mass eigenstates.
On the other hand, these matrices do not appear from sfermion components, because 
we approximate that sfermions are in the mass eigenstate basis.
In the present model, there are only three R-parity violating trilinear interactions allowed by the 
family symmetry, and baryon number violating terms $\lambda''_{ijk}U^c_i D^c_j D^c_k$ 
are forbidden by the symmetry.  
It should be compared with the MSSM case in which there are 
45 trilinear couplings. 
These interactions can generate a lot of new processes which have not been observed yet 
such as lepton flavor violating (LFV) processes, 
or new contributions to already observed processes.
Many authors have studied phenomenology of R-parity violation and  
obtained constraints on each coupling constant corresponding to each process 
in the MSSM case.
\footnote{See Ref. \cite{rpreview,allanach} and references therein.}

In this section, we obtain constraints on the coupling constants $\lambda, \lambda'_{1,2}$ 
at the weak scale.
Since the various new processes generated by the interactions
$W_{R\hspace{-2mm}/}$ depend only on the three coupling constants, we can predict ratios 
of new processes independent of $\lambda$s.  
We will also find the ratios of the LFV processes in this section.
As mentioned before, we assume that the R-parity violating couplings $\lambda$ and $\lambda'_{1,2}$ are real and positive \footnote{CP violation induced by the R-parity violating trilinear couplings 
in the soft SUSY breaking sector has been studied in Ref. \cite{abel}.}.

\subsection{Constraint on $\lambda$}
In this subsection, we consider the constraint on $\lambda L_3 L_I E^c_I$ operator.
The most stringent constraint on $\lambda$ is obtained from both $\mu \to eee$ 
and neutrinoless double beta decay, both the processes give similar bound.

\begin{figure}[t]
\unitlength=1mm
\begin{picture}(60,30)
\includegraphics[width=7cm]{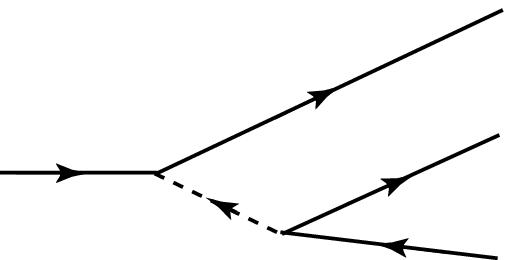}
\put(-50,16){$\lambda$}
\put(-65,16){$\ell_m^-$}
\put(-8,10){$\ell^-_j$}
\put(-8,28){$\ell^-_i$}
\put(-8,-5){$\ell^+_k$}
\put(-44,4){$\tilde \nu$}
\put(-31,-2){$\lambda$}
\end{picture}
\hspace{2cm}
\begin{picture}(60,30)
\includegraphics[width=7cm]{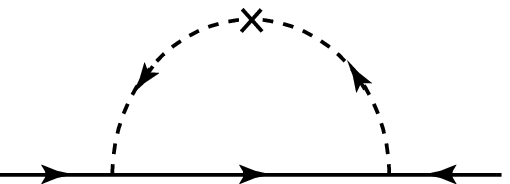}
\put(-39,27){${\bf m}^2_{\tilde e LR}$}
\put(-57,15){$\tilde e_L$}
\put(-16,15){$\tilde e_R$}
\put(-8,5){$\nu_j$}
\put(-66,5){$\nu_i$}
\put(-17,-3){$ \lambda$}
\put(-56,-3){$\lambda$}
\put(-37,-3){$e$}
\end{picture}
\vspace{1.0cm}
\caption{R-parity violating contributions to the decay processes 
$\ell_m^- \to \ell_i^- \ell_j^- \ell_k^+$ (left) and the neutrino mass (right).}
\label{lle}
\end{figure}

First we show the bound on $\lambda$ from $\mu \to eee$ process \cite{hall, lfv, hinchliffe,choudhury,gouvea}.
The decay process $\ell_m^- \to \ell_i^- \ell_j^- \ell_k^+$ is generated by tree-level 
$t$- and $u$- channel sneutrino 
exchange (FIG. \ref{lle}), and its effective Lagrangian is given by
\be
{\cal L}_{eff}=\lambda^2 A_L \left(\bar \ell_iP_L \ell_m\right) \left( \bar \ell_jP_R \ell_k\right)
+\lambda^2 A_R\left(\bar \ell_iP_R \ell_m\right) \left( \bar \ell_jP_L \ell_k\right)+(i \leftrightarrow j),
\label{mueee}
\ee
where the coefficients are 
\be
A_L&=& \frac{1}{m_{\tilde \ell 1L}^2}(U_{eR}^\dag)_{iJ}(U_{eR})_{Jk}
(U_{eL}^\dag)_{j3}(U_{eL})_{3m}+\frac{1}{m_{\tilde \ell 3L}^2}(U_{eR}^\dag)_{iJ}(U_{eL})_{Jm}
(U_{eL}^\dag)_{jK}(U_{eR})_{Kk}, \\
A_R&=&\frac{1}{m_{\tilde \ell 1L}^2}(U_{eR}^\dag)_{jJ}(U_{eR})_{Jm}
(U_{eL}^\dag)_{i3}(U_{eL})_{3k}+\frac{1}{m_{\tilde \ell 3L}^2}(U_{eR}^\dag)_{jJ}(U_{eL})_{Jk}
(U_{eL}^\dag)_{iK}(U_{eR})_{Km},
\label{mueeeamp}
\ee
with mixing matrices in Eq.(\ref{UeL}), (\ref{UeR}). 
In our approximation, sneutrino mass is the same as that of the left-handed slepton defined in 
Eq.(\ref{scalarmass}).
From this Lagrangian, the branching ratio of $\mu \to eee$ is given by, 
\be
BR(\mu \to eee)=\frac{4 \lambda^4}{64 G_F^2}\left[ |A_L|^2+|A_R|^2\right]
BR(\mu \to e \bar \nu_e \nu_{\mu}).
\label{mueeebr}
\ee
The requirement that this branching ratio should not exceed the experimental bound 
$BR(\mu\to eee)^{exp}<1.0 \times 10^{-12}$ provides a constraint on $\lambda$
\be
\lambda<1.4 \times 10^{-2} \left( \frac{m_{\tilde \ell L}}{100 \GeV}\right),
\label{lam1}
\ee
where we have assumed $m_{\tilde \ell 1L}=m_{\tilde \ell 3L}\equiv m_{\tilde \ell L}$ 
in order to forbid the contribution to FCNC processes from the soft scalar mass terms.

Next we show the constraint from neutrinoless double beta decay. 
The coupling $\lambda$ 
can induce radiative neutrino mass at the one loop level shown in 
FIG. \ref{lle} \cite{hall,rnumass}:

\be
\left( {\cal M}_{\nu}\right)_{ij}&=&-2 \frac{\lambda^2}{(4 \pi)^2}m_{ek}
\left[ (U_{eL})_{3j}(U_{eL})_{Jk}-(U_{eL})_{3k}(U_{eL})_{Jj}\right] \times \nn \\
&\times &\left[ (U_{eR}^\dag)_{kI}(U_{eL})_{3i}({\bf m}^2_{\tilde e LR})_{IJ}
F(m^2_{ek},m^2_{\tilde e 1R},m^2_{\tilde \ell 1L})
-(U_{eR}^\dag)_{kK}(U_{eL})_{Ki}({\bf m}^2_{\tilde e LR})_{3J}
F(m^2_{ek},m^2_{\tilde e 1R},m^2_{\tilde \ell 3L})\right] \nn \\
&+& (i \leftrightarrow j),
\ee
where the loop function $F$ is
\be
F(x,y,z)&=&\frac{x \ln x}{(x-y)(x-z)}+\frac{y \ln y}{(y-z)(y-x)}+\frac{z \ln z}{(z-x)(z-y)} \nn \\
&\rightarrow& \frac{\ln (y/z)}{y-z},~~\mbox{as}~~ x \to 0. 
\ee
We assume that the trilinear terms in the soft SUSY breaking sector are completely 
aligned with the Yukawa matrices, that is,
\be
{\bf m}^2_{\tilde e LR} \simeq {\bf m}^e \tilde A_e,
\ee
in which $\tilde A_e$ stands for the sum of the A-terms in Eq.(\ref{Aterm}) and $\mu$-terms 
which are not explicitly shown, and 
${\bf m}^e$ is the mass matrix of charged leptons Eq.(\ref{me}). 
The requirement that the effective neutrino mass $({\cal M}_{\nu})_{ee}$ does not exceed a 
constraint from neutrinoless double beta decay 
provides a bound on $\lambda$: 
\be
\lambda<1.1 \times 10^{-2}\left( \frac{({\cal M}_{\nu})_{ee}^{exp}}{0.35{\mbox eV}}\right)^{1/2}
\left( \frac{100 \GeV}{\tilde M}\right)^{1/2},~~~
\tilde M=\tilde A_e \frac{(100 \GeV)^2}{{m^2_{\tilde e R}-m^2_{\tilde \ell L}}}
\ln \frac{m^2_{\tilde e R}}{m^2_{\tilde \ell L}}.
\label{lam2}
\ee
This is the most stringent constraint on $\lambda$ in the present model: 
other processes give weaker bounds. For example, a constraint from 
$\mu \to e \gamma$ is \cite{chaichian}
\be
\lambda< 0.22, 
\ee
when slepton masses are $100$GeV.

\subsection{Constraints on $\lambda_{1}'$ and $\lambda_{2}'$}
Constraints on $\lambda_{1,2}'$ are obtained from neutral meson mixings 
\cite{mesonmixing,choudhury,carlos,bhattacharyya,saha,kundu,wang}, and 
on the products $\lambda \lambda_{1}'$ and $\lambda \lambda_{2}'$ 
from leptonic decays of neutral mesons 
\cite{choudhury,bhattacharyya,grossman,jang,dreiner}.
Since both processes are generated at tree level, these give the most stringent bounds on 
$\lambda'_{1,2}$. Although $\mu -e$ conversion in nuclei \cite{gouvea,kim} is also generated at tree level by $\lambda'_{1,2}$, bounds from this process are weaker than those 
from neutral meson system. Therefore, we show the calculations only of the neutral meson system.

\begin{figure}[t]
\unitlength=1mm
\begin{picture}(70,30)
\includegraphics[width=6cm]{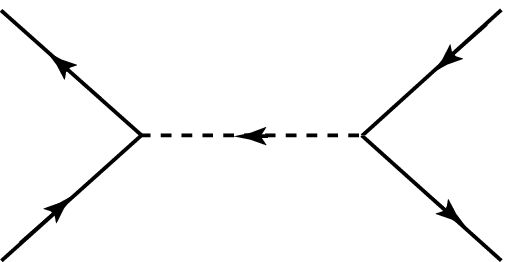}
\put(-32,18){$\tilde \nu$}
\put(-43,19){$\lambda'$}
\put(-25,19){$\lambda'(\lambda)$}
\put(1,0){$d_j(\ell_f)$}
\put(1,30){$\bar d_i(\bar \ell_n)$}
\put(-64,-3){$d_i$}
\put(-64,30){$\bar d_j $}
\end{picture}
\hspace{1cm}
\begin{picture}(60,30)
\includegraphics[width=6cm]{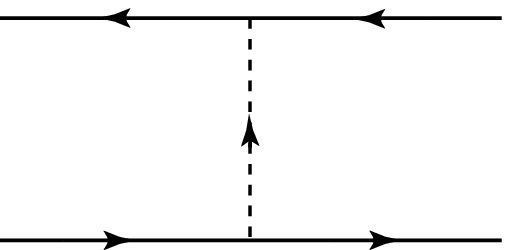}
\put(-31,30){$\lambda'$}
\put(-31,-4){$\lambda'$}
\put(-28,14){$\tilde \nu(\tilde u)$}
\put(1,0){$d_j(\ell_f)$}
\put(1,28){$\bar d_i(\bar \ell_n)$}
\put(-64,0){$d_i$}
\put(-64,28){$\bar d_j $}
\end{picture}
\vspace{1.0cm}
\caption{R-parity violating contributions to 
the neutral meson mixings (or, leptonic dcays of the 
neutral mesons) 
in the $s$- and $t$- channels. The quantities in parenthesis represent the case 
of leptonic decays.}
\label{st}
\end{figure}

The neutral meson mixing is generated at the tree level through the exchange of a sneutrino 
in both $s$- and $t$- channels (FIG. \ref{st}).
For $K^0-\bar K^0$ mixing, the effective Hamiltonian is obtained as
\be
{\cal H}_{eff}=\frac{\Lambda'_{I21}\Lambda'^{*}_{I12}}{m^2_{\tilde \ell 1L}}
(\bar d_R s_L)(\bar d_L s_R),
\label{k0mixham}
\ee
where
\be
\Lambda'_{Ijk}=\lambda'_1(U_{dR}^\dag)_{jJ}(U_{dL})_{3k}(i \sigma^2)_{IJ}
+\lambda'_2 (U_{dR}^\dag)_{j3}(U_{dL})_{Jk}(\sigma^1)_{IJ}.
\label{k0mixamp}
\ee
We require that these additional contributions to the mass difference of neutral K meson are 
smaller than its experimental value:
\be
\Delta m_{K^0}^{R\hspace{-2mm}/}=\frac{|\Lambda'_{I21}\Lambda'^{*}_{I12}|}{2 m^2_{\tilde \ell 1L}}
S_{K^0}m_{K^0}f_K^2 B_4(m_{K^0})<\Delta m_{K^0}^{exp},
\ee
where $m_{K^0}$ and $f_K$ denote mass and the decay constant of the neutral K meson, and 
\be
S_{K^0}=\left( \frac{m_{K^0}}{m_s(\mu)+m_d(\mu)}\right)^2,~~B_4(m_{K^0})=1.03,~~
\Delta m_{K^0}^{exp}=\left(0.5292 \pm 0.0009\right) \times 10^{-2} \mbox{ps}^{-1} 
\ee
with $\mu=2$ GeV \cite{kundu}.
Then \cite{choudhury}, 
\be
|\Lambda'_{I21}\Lambda'^{*}_{I12}|<4.5 \times 10^{-9} 
\left( \frac{m_{\tilde \ell L}}{100 \GeV}\right)^2. 
\ee
The assumptions $\theta^d_1-\theta^d_3=0$ and $\lambda'_1=\lambda'_2$
lead to the most stringent constraints on $\lambda'_{1,2}$:
\be
\lambda'_1=\lambda'_2<3.1 \times 10^{-3}\left( \frac{m_{\tilde \ell L}}{100 \GeV}\right).
\label{lamp12}
\ee

Next, we consider the leptonic decays of the neutral K mesons: 
$K_{L,S}\to e^- e^+, \mu^- \mu^+$ or $ e^\mp \mu^\pm$. 
At the quark level, these processes are interpreted as a transformation of 
a down-type quark-antiquark pair ($d_i$ and $\bar d_j$) into a 
charged lepton-antilepton pair ($\ell_f$ and $\bar \ell_n$), which are 
shown in FIG. \ref{st}. 
The processes in the $s$- and $t$- channels are mediated by 
sneutrino and $u$-squark, respectively. 
In general, the effective Lagrangian for 
the process $d_i \bar d_j \to \ell_f \bar \ell_n$ is written as \cite{dreiner}
\be
{\cal L}_{eff}&=&-\frac{1}{2}{\cal A}_{ijfn}(\bar d_{jR}\gamma^{\mu}d_{iR})
(\bar \ell_{fL}\gamma_{\mu}\ell_{nL})\nn \\
&+&\frac{1}{2}\frac{m_{di}+m_{dj}}{m_M^2}
\left[ (\bar d_{jL}d_{iR})(\bar \ell_{fR}\ell _{nL}){\cal B}_{ijfn}
+(\bar d_{jR}d_{iL})(\bar \ell_{fL}\ell _{nR}){\cal B}_{jinf}^*\right],
\label{kleff}
\ee
where each coefficient is defined as
\be
{\cal A}_{ijfn}&=&\frac{\lambda'^2_1}{m^2_{\tilde q 3L}}\left\{ (U_{eL}^\dag)_{fI}(i \sigma^2)_{IJ}
(U_{dR})_{Ji}\right\}\left\{ (U_{eL})_{Ln}(i \sigma^2)_{LK}(U_{dR}^\dag)_{jK}\right\} \nn \\
&+&\frac{\lambda'^2_2}{m^2_{\tilde q 1L}}(U_{dR})_{3i}(U_{dR}^\dag)_{j3}(U_{eL}^\dag)_{fI}
(U_{eL})_{In},\label{klampa} \\
{\cal B}_{ijfn}&=&-2 \frac{m_{M}^2}{m_{di}+m_{dj}}\frac{\lambda}{m^2_{\tilde \ell 1L}}
(U_{eR}^\dag)_{fK}(U_{eL})_{3n}\left[ \lambda'_1(U_{dR})_{Ji}(U_{dL}^\dag)_{j3}(i \sigma^2)_{KJ}
+\lambda'_2 (U_{dR})_{3i}(U_{dL}^\dag)_{jJ}(\sigma^1)_{KJ}
\right],\nn \\
\label{klampb}
\ee
and $m_M$ stands for the mass of the meson composed of quark($d_i$)-antiquark($\bar d_j$) pair 
with mass $m_{di}$ and $m_{dj}$.
From the above effective Lagrangian, the decay rate of $d_i \bar d_j \to \ell_f \bar \ell_n$ is 
given by
\be
\Gamma(d_i \bar d_j \to \ell_f \bar \ell_n)=
\frac{f_M^2}{256 \pi m_M^3}{\cal C}_{ijfn}\sqrt{\left\{ m_M^2-(m_f-m_n)^2\right\}
\left\{ m_M^2-(m_f+m_n)^2\right\}},
\label{kldr}
\ee
where 
\be
{\cal C}_{ijfn}&=&\left(m_M^2-m_f^2\right)\bigl| {\cal A}_{ijfn}m_f+{\cal B}_{ijfn}\bigr|^2
+\left(m_M^2-m_n^2 \right)\bigl| {\cal A}_{ijfn}m_n+{\cal B}_{jinf}^*\bigr|^2 \nn \\
&-&\bigl| {\cal B}_{ijfn}m_n-{\cal B}_{jinf}^*m_f \bigr|^2
+m_f m_n \left[ \bigl| {\cal B}_{ijfn}+{\cal B}_{jinf}^*\bigr|^2
-\bigl| {\cal A}_{ijfn}m_n-{\cal B}_{ijfn}\bigr|^2 \right. \nn \\
&-&\left. \bigl| {\cal A}_{ijfn}m_f-{\cal B}_{jinf}^*\bigr|^2
+\bigl| (m_f+m_n){\cal A}_{ijfn}\bigr|^2 \right],
\label{klampc}
\ee
and $f_M$ is the decay constant of the meson under consideration and 
$m_{f,n}$ are the corresponding charged lepton masses.
In the case of $K_L$ decay, ${\cal A}_{ijfn}$ should be replaced to 
$({\cal A}_{ijfn}-{\cal A}_{jifn})/\sqrt{2}$ with $i=1,j=2$
, and similar for ${\cal B}_{ijfn}$ and ${\cal B}_{jinf}^*$. 

We require that branching ratios should not exceed the experimental bound
$BR(K_L \to \mu^\mp e^\pm)^{exp}<4.7 \times 10^{-12}$ and the experimantal value 
$BR(K_L \to e^- e^+)^{exp}=9 \times 10^{-12}$, therefore we obtain
\be
\lambda \lambda'_1 &<& 5.4 \times 10^{-7} \left( \frac{m_{\tilde \ell L}}{100 \GeV}\right)^2
 \label{lamlamp1}
\ee
from $K_L \to \mu^{\mp}e^{\pm}$, and 
\be
\lambda \lambda'_2 &<& 1.1 \times 10^{-8} \left( \frac{m_{\tilde \ell L}}{100 \GeV}\right)^2
\label{lamlamp2}
\ee
from $K_L \to e^- e^+$.
The contributions including up type squarks, that is, ${\cal A}_{ijfn}$, are negligible in
 $K_L \to \mu^{\mp}e^{\pm}$ because these terms only provide weaker constraints, and vanish in
$K_L \to e^- e^+$ because of the replacement mentioned above. 

For comparison, bounds on $\lambda \lambda'_{1,2}$ from $\mu-e$ conversion in nuclei are 
$\lambda \lambda'_1<10^{-6}$ and $\lambda \lambda'_2<10^{-7}$, which are weaker than 
Eqs. (\ref{lamlamp1}) and (\ref{lamlamp2}). 

\subsection{Predictions for Lepton Flavor Violating processes}
\begin{figure}[t]
\unitlength=1mm
\begin{center}
\begin{picture}(50,31)
\includegraphics[width=7cm]{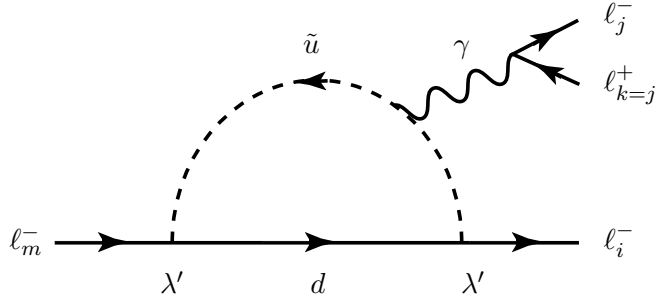}
\put(-76,1){$\ell^-_m$}
\put(3,22){$\ell_{k=j}^+$}
\put(3,31){$\ell_j^-$}
\put(3,1){$\ell^-_i$}
\put(-17,27){$\gamma$}
\put(-37,27){$\tilde u$}
\put(-36,-5){$d$}
\put(-56,-5){$\lambda'$}
\put(-16,-5){$\lambda'$}
\end{picture}
\end{center}
\caption{The photon penguin contributions to the decays 
$\ell_m^- \to \ell_i^- \ell_j^- \ell_{k=j}^+$. These contributions can be negligible 
compared to the tree level processes induced by the coupling $\lambda$.}
\label{penguin}
\end{figure}

Although many processes can be generated by the R-parity violating interactions, 
we focus on the LFV decays $\ell_m^- \to \ell_i^- \ell_j^- \ell_k^+$, where $m=\mu$ or $\tau$,  
in this subsection. As mentioned in the subsection {\bf A.}, the operator $\lambda L_3 L_I E_I^c$ 
generates the decays $\ell_m^- \to \ell_i^- \ell_j^- \ell_k^+$ at tree level when $\lambda \neq 0$. 
The other two operators in Eq.(\ref{rparity}), $\lambda'_{1,2}LQD^c$, also generate the similar decay processes
at one loop level through photon penguin diagrams shown in FIG. \ref{penguin}, 
but we found that the bounds on 
$\lambda'_{1,2}$ are stronger than that on $\lambda$ in the previous subsections. 
So we neglect contributions from $\lambda'_{1,2}$ operators to the decays
 $\ell_m^- \to \ell_i^- \ell_j^- \ell_k^+$.   
Moreover, flavor changing $Z$ boson decay $Z \to \ell^-_i \ell^+_j$ induced by the R-parity 
violating bilinear terms can contribute to  $\ell_m^- \to \ell_i^- \ell_j^- \ell_k^+$ processes. 
Since branching ratios of these decays are propotional to $\sin^2 \xi$, their effects are also negligible
\cite{bisset}. Besides these R-parity violating contributions, there are two other contributions to these 
processes by Higgs bosons. Since the charged leptons couple to the neutral Higgs bosons, 
these Yukawa interactions generate LFV processes at one loop level \cite{kolda}. 
However, these effects are enhanced only when $\tan \beta$ is large. 
So, we assume that these are negligible because $\tan \beta$ is small enough. 
Moreover, since there are three generations of both up and down type Higgs doublet in this model, 
LFV processes mediated by the neutral Higgs bosons are generated at tree level.
However, the branching ratio of the $\mu \to eee$ from these effects is $BR \sim 10^{-16}$ 
because of the smallness of the Yukawa couplings 
when the neutral Higgs boson mass is $100 \GeV$. So, these contributions can also be negligible 
compared to those from $\lambda LLE^c$ couplings unless $\lambda<10^{-3}$. 
Therefore, we can 
approximate that $\ell_m^- \to \ell_i^- \ell_j^- \ell_k^+$ processes are induced at tree level 
only by 
$\lambda$. In this approximation, the ratios of these processes are independent of $\lambda$, 
but depend on the mixing matrices $U_{eL(R)}$ which reflect the flavor structure of the 
model. Therefore we find some predictions of LFV decays $\ell_m^- \to \ell_i^- \ell_j^- \ell_k^+$ in our model.

From the branching ratio Eq.(\ref{mueeebr}), we can easily find the ratios of processes in the 
approximation that all scalar masses are equal:\footnote{From the conditions to suppress 
$\mu \to e+\gamma$ process from the scalar mass terms, slepton masses are required to be degenerated with mass differences of order $10^{-1}$\cite{kubo2,kajiyama}. }
\be
\frac{BR(\tau \to eee)}{BR(\tau \to \mu \mu \mu)}
&\simeq& \frac{4 \epsilon_{\mu}^2}{1 +\epsilon_{\mu}^2}=0.014, \\
\frac{BR(\tau \to \mu\mu e)}{BR(\tau \to \mu \mu \mu)}
&\simeq& \frac{1-\epsilon_{\mu}^2+2 \epsilon_e^2}{1+\epsilon_{\mu}^2-2 \epsilon_e^2}
=0.99, \\
\frac{BR(\mu \to eee)}{BR(\tau \to eee)}
&\simeq& \frac{\tau_{\mu}}{\tau_{\tau}}\epsilon_{\mu}^5 
\frac{\epsilon_e^2}{2 \epsilon_{\mu}^2+\epsilon_e^2}=0.0093,
\ee
where small parameters $\epsilon_{e,\mu}$ are given in Eq.(\ref{epsilon}) and $\tau_{\mu}(\tau_{\tau})$ stand for the lifetime of the $\mu (\tau)$ lepton. 
Also, $BR(\tau \to \mu \mu e)$ means $BR(\tau^- \to \mu^- \mu^- e^+)$, 
and similar for the other processes.
One can obtain the ratios of 
other combinations from the branching ratios listed below:
\be
BR(\mu \to eee)&\propto& \tau_{\mu} \epsilon_e^2 \left[ 
2(1-2 \epsilon_{\mu}^2)m^{-4}_{\tilde \ell 1L}
+\frac 12 m^{-4}_{\tilde \ell 3L}
-2(1-\epsilon^2_{\mu})m^{-2}_{\tilde \ell 1L}m^{-2}_{\tilde \ell 3L}\right], \\
BR(\tau \to eee)&\propto& \tau_{\tau}\left[ 
\epsilon^2_{\mu}\left( 1-\epsilon^2_{\mu}-4\epsilon^2_e\right)m^{-4}_{\tilde \ell 1L}
+\frac 12 \epsilon^2_e m^{-4}_{\tilde \ell 3L}\right],\\
BR(\tau \to \mu \mu \mu)&\propto&\tau_{\tau}\left[ 
\frac 14 \left(1+\epsilon^2_{\mu}-2 \epsilon^2_e-4 \epsilon^2_e \epsilon^2_{\mu} \right)
m^{-4}_{\tilde \ell 3L}
-2 \epsilon^2_e \epsilon^2_{\mu}m^{-2}_{\tilde \ell 1L}m^{-2}_{\tilde \ell 3L}\right],\\
BR(\tau \to ee\mu)&\propto&\tau_{\tau}\epsilon^2_e \left[ 
2 \epsilon^2_{\mu}m^{-4}_{\tilde \ell 1L}
+(\frac 12-\epsilon^2_{\mu})m^{-4}_{\tilde \ell 3L}
-2 \epsilon^2_{\mu}m^{-2}_{\tilde \ell 1L}m^{-2}_{\tilde \ell 3L}\right],\\
BR(\tau \to \mu \mu e)&\propto& \tau_{\tau}\frac 14 \left(1-\epsilon^2_{\mu}
+2 \epsilon^2_e-4  \epsilon^2_e \epsilon^2_{\mu} +\frac 14 \epsilon^4_{\mu}  \right)
m^{-4}_{\tilde \ell 3L},\\
BR(\tau \to \mu ee)&\propto&\tau_{\tau}\left[ 
\epsilon^2_e \epsilon^2_{\mu}m^{-4}_{\tilde \ell 1L}
+\frac 18\left( 1-2 \epsilon^2_{\mu}+6\epsilon^2_e \epsilon^2_{\mu}
+\frac 32 \epsilon^4_{\mu}\right)m^{-4}_{\tilde \ell 3L}  \right],\\
BR(\tau \to \mu e \mu)&\propto& \tau_{\tau}\left[ 
\frac 18 \left( 1-4 \epsilon^2_e+6 \epsilon^2_e \epsilon^2_{\mu}
-\frac 34 \epsilon^4_{\mu}\right)m^{-4}_{\tilde \ell 3L}
+2 \epsilon^2_e \epsilon^2_{\mu}m^{-2}_{\tilde \ell 1L}m^{-2}_{\tilde \ell 3L}\right],
\ee
where the common factor is not shown explicitly.


\section{Conclusion}

We have considered the properties of R-parity violating operators 
in a SUSY model with non-Abelian discrete $Q_6$ family symmetry. 
The family symmetry can reduce the number of parameters in the Yukawa 
sector, and explain the fermion masses and mixings between generations. 
It can also reduce the number of R-parity violating couplings 
and determine the form of those. Only three trilinear couplings are allowed, and 
the baryon number violating operators are forbidden by the symmetry in our model. 
We derived upper bounds on these couplings: $\lambda<O(10^{-2})$, 
$\lambda'_{1,2}<O(10^{-4})$, and obtained the predictions on the ratios of the LFV decays 
$\ell_m^- \to \ell_i^- \ell_j^- \ell_k^+$ which do not depend unknown parameters. 
The results reflect the properties of the family symmetry because these predictions contain 
the mixing matrices of the charged lepton sector which is written by masses of the charged leptons. 
Our predictions can be testable at future experiments because the superB factory \cite{kekb} or LHC 
\cite{tau3mu} will have the sensitivity $BR\sim 10^{-(8\div 9)}$ for LFV $\tau$ decays.

\vspace{0.5cm}
\noindent
{\large \bf Acknowledgments}\\
The author would like to thank A. Hektor, K. Kannike, N. Kifune, J. Kubo and M. Raidal for useful discussions.
This work is supported by the ESF grant No. 6190 and postdoc
contract 01-JD/06.                                                                           
\bibliographystyle{unsrt}

\begin{thebibliography}{99}
\bi{babukubo}
K. S. Babu and J. Kubo, Phys. Rev. {\bf D71}, 056006 (2005).

\bi{kajiyama}
E. Itou, Y. Kajiyama and J. Kubo, Nucl. Phys. {\bf B743}, 74 (2006). 

\bi{kubo}
J.~Kubo, A.~Mondrag\'on, M.~Mondrag\'on and
E.~Rodr\' iguez-J\' auregui,  Prog. Theor. Phys.

{\bf 109}, 795 (2003); Erratum-ibid.{\bf 114}, 287 (2005);
J.~Kubo, Phys. Lett. {\bf B578}, 156 (2004);
Erratum-ibid. {\bf B619}, 387 (2005); 
Y. Kajiyama, J. Kubo and H. Okada, Phys. Rev. {\bf D75}, 033001 (2007).

\bi{kubo2}
T. Kobayashi, J. Kubo and H. Terao, Phys. Lett. {\bf B568}, 83 (2003);
K-Y. Choi, Y. Kajiyama, H. M. Lee and J. Kubo, Phys. Rev. {\bf D70}, 055004 (2004).

\bi{frampton}
P. H. Frampton and T. W. Kephart,
Int. J. Mod. Phys. {\bf A10}, 4689 (1995); 
Phys. Rev. {\bf D64}, 086007 (2001).

\bi{weinberg}S.~Weinberg,
in {\em Transactions of the New York Academy of Sciences}, 
Series II, Vol. 38, 185 (1977);
F.~Wilczek and A.~Zee,
Phys. Rev. Lett. {\bf 42}, 421 (1979).

\bi{fritzsch1}H.~Fritzsch,
Phys. Lett. {\bf B 73}, 317 (1978);  Nucl. Phys. {\bf B155}, 189 (1979). 

\bi{pdg}
W. M. Yao {\em et al}. [Particle Data Group],
J. Phys. G {\bf 33}, 1 (2006); 
CKMfitter [http://www.slac.stanford.edu/xorg/ckmfitter/].

\bi{deltams}
A. Abulencia {\em et al}. [CDF collaboration], 
Phys. Rev. Lett. {\bf 97}, 062003 (2006).

\bibitem{seesaw} P.~Minkowski, Phys. Lett. {\bf B67}, 421 (1977).
T. Yanagida,
 in {\em Proc, of the Workshop on the unified Theory
 and Baryon Number in the universe,} 
 ed. O. Sawada and A. Sugamoto,
 (KEK report 79-18, 1979);
 M. Gell-Mann, P. Ramond and R. Slansky, 
 in {\em Supergravity},
 ed P. van Nieuwenhuizen and d.Z. Freedman,
 (North Holland, Amsterdam, 1979);
 R.~N.~Mohapatra and G.~Senjanovic,
Phys.  Rev.  Lett.   {\bf 44},  912 (1980).

\bi{farrar}
G. R. Farrar and P. Fayet, 
Phys. Lett. {\bf B76}, 575 (1978).

\bi{rpreview}
G. Bhattacharyya,
Nucl. Phys. Proc. Suppl. {\bf 52A}, 83 (1997);
H. K. Dreiner,  hep-ph/9707435;
P. Roy, hep-ph/9712520; 
M. Chemtob,
Prog. Part. Nucl. Phys. {\bf 54}, 71 (2005);
R. Barbier {\em et al}. 
Phys. Rept. {\bf 420}, 1 (2005).

\bi{ibanez}
L. E. Ibanez and G. G. Ross,
Nucl. Phys. {\bf B368},3 (1992).

\bi{hall}
L. J. Hall and M. Suzuki,
Nucl. Phys. {\bf B231}, 419 (1984);

\bi{lfv}
S. Dawson, 
Nucl. Phys. {\bf B261}, 297 (1985);
V. Barger, G. F. Giudice and T. Han,
Phys. Rev. {\bf D40}, 2987 (1989).

\bi{hinchliffe}
I. Hinchliffe and T. Kaeding, 
Phys. Rev. {\bf D47}, 279 (1993).    

\bi{gouvea}
A. de Gouv\^ea, S. Lola and K. Tobe,
Phys. Rev. {\bf D63}, 035004 (2001). 

\bi{kim}
J. E. Kim, P. Ko and D-G. Lee, 
Phys. Rev. {\bf D56}, 100 (1997); 
K. Huitu, J. Maalampi, M. Raidal and A. Santamaria, 
Phys. Lett. {\bf B430}, 355 (1998). 

\bi{chaichian}
M. Chaichian and K. Huitu,
Phys. Lett. {\bf B384}, 157 (1996);
B. de Carlos and P. L. White, 
Phys. Rev. {\bf D54}, 3427 (1996). 

\bi{rnumass}
R. N. Mohapatra, 
Phys. Rev. {\bf D34}, 3457 (1986); 
S. Dimopoulos and L. J. Hall, 
Phys. Lett. {\bf B207}, 210 (1987);
K. S. Babu and R. N. Mohapatra, 
Phys. Rev. Lett. {\bf 64}, 1705 (1990);
G. Bhattacharyya and H. V. Klapdor-Kleingrothaus and 
H. P\"as, Phys. Lett. {\bf B463}, 77 (1999).

\bi{joshipura}
A. S. Joshipura and M. Nowakowski, 
Phys. Rev. {\bf D51}, 2421 (1995);
R. Hempfling, 
Nucl. Phys. {\bf B478}, 3 (1996).

\bi{banks}
T. Banks, Y. Grossman, E. Nardi and Y. Nir,
Phys. Rev. {\bf D52}, 5319 (1995);
P. Bin\'etruy, E. Dudas, S. Lavignac and C. A. Savoy, 
Phys. Lett. {\bf B422}, 171 (1998).  

\bi{bisset}
M. Bisset, O. C. W. Kong, C. Macesanu and L. H. Orr,
Phys. Lett. {\bf B430}, 274 (1998).                              

\bi{mesonmixing}
R. Barbieri and A. Masiero,
Nucl. Phys. {\bf B267}, 679 (1986).

\bi{carlos}
B. de Carlos and P. L. White,
Phys. Rev. {\bf D55}, 4222 (1997).

\bi{bhattacharyya}
G. Bhattacharyya and A. Raychaudhuri,
Phys. Rev. {\bf D57}, 3837 (1998).

\bi{saha}
J. P. Saha and A. Kundu,
Phys. Rev. {\bf D69}, 016004 (2004).

\bi{kundu}
A. Kundu and J. P. Saha,
Phys. Rev. {\bf D70}, 096002 (2004). 

\bi{wang}
R. M. Wang, G. R. Lu, E. K. Wang and Y. D. Yang,
hep-ph/0609276.

\bi{grossman}
Y. Grossman, Z. Ligeti and E. Nardi,
Phys. Rev. {\bf D55}, 2768 (1997).

\bi{jang}
J-H. Jang, J. K. Kim and J. S. Lee, 
Phys. Rev. {\bf D55}, 7296 (1997);
S. Baek and Y. Kim, 
Phys. Rev. {\bf D66}, 077701 (1999); 
J. P. Saha and A. Kundu, 
Phys. Rev. {\bf D66}, 054021 (2002).

\bi{dreiner}
H. Dreiner, G. Polesello and M. Thormeier, 
Phys. Rev. {\bf D65}, 115006 (2002).

\bi{benhamo}
V. Ben-Hamo and Y. Nir, 
Phys. Lett. {\bf B339}, 77 (1994).

\bi{pdecay}
J. L. Goity and M. Sher, 
Phys. Lett. {\bf B346}, 69 (1995); 
Erratum-ibid, {\bf B385}, 500 (1996);
C. E. Carlson, P. Roy and M. Sher, 
Phys. Lett. {\bf B357}, 99 (1995); 
J. Hisano, 
hep-ph/0004266; 
F. Vissani, 
Phys. Rev. {\bf D52}, 4245 (1995); 
A. Y. Smirnov and F. Vissani, 
Phys. Lett. {\bf B380}, 317 (1996); 
G. Bhattacharyya and P. B. Pal, 
Phys. Rev. {\bf D59}, 097701 (1999); 
Phys. Lett. {\bf B439}, 81 (1998).

\bi{choudhury}
D. Choudhury and P. Roy, 
Phys. Lett. {\bf B378},153 (1996);
K. Agashe and M. Graesser, 
Phys. Rev. {\bf D54}, 4445 (1996).

\bi{abelian}
A. H. Chamseddine and H. Dreiner, 
Nucl. Phys. {\bf B458}, 65 (1996);
P. Bin\'etruy, S. Lavignac and P. Ramond, 
Nucl. Phys. {\bf B477}, 353 (1996); 
E. J. Chun and A. Lukas, 
Phys,. Lett. {\bf B387}, 99 (1996); 
J. Ellis, S. Lola and G. G. Ross, 
Nucl. Phys. {\bf 526}, 115 (1998); 
G. Eyal and Y. Nir, 
JHEP. {\bf 9906}, 024 (1999);
A. S. Joshipura, R. Vaidya and S. K. Vempati, 
Phys. Rev. {\bf D62}, 093020 (2000).

\bi{nonabelian}
G. Bhattacharyya, 
Phys. Rev. {\bf D57}, 3944 (1998).

\bi{allanach}
B. C. Allanach, A. Dedes and H. K. Dreiner, 
Phys. Rev. {\bf D60}, 075014 (1999).

\bi{abel}
S. A. Abel, 
Phys. Lett. {\bf B410}, 173 (1997).

\bi{kolda}
K. S. Babu and C. Kolda, 
Phys. Rev. Lett. {\bf 89}, 241802 (2002).

\bi{kekb}
A. G. Akeroyd {\em et al}. [SuperKEKB Physics Working Group],
KEK report {\bf 04-4}, [arXiv:hep-ex/0406071].

\bi{tau3mu}
R. Santinelli and M. Biasini, 
CMS-note/2002-037; 
N. G. Unel, 
hep-ex/0505030.

\end{thebibliography}

\end{document}